\RequirePackage{lineno}
\documentclass[aps,prl,twocolumn,superscriptaddress]{revtex4}

\usepackage{graphicx}
\usepackage{bm}
\usepackage{amsmath}
\usepackage{amssymb}
\usepackage{hyperref}
\usepackage{siunitx}

\begin{document}
\title{Gear Junctions between Chiral Boron Nitride Nanotubes}

\author{Zhao Wang}
\email{zw@gxu.edu.cn}

\affiliation{Department of Physics, Guangxi University, 530004 Nanning, China; and Institute of Materials Chemistry, TU Wien, A-1060 Vienna, Austria}

\begin{abstract}
A gear effect is demonstrated at parallel and cross junctions between boron nitride nanotubes (BNNTs) via atomistic simulations. The atoms of neighboring BNNTs are meshed together at the junctions like gear teeth through long-range non-covalent interaction, which are shown to be able to transmit motion and power. The sliding motion of a BNNT can be spontaneously translated to rotating motion of an adjoining one or viceversa at a well-defined speed ratio. The transmittable motion and force strongly depend on the helical lattice structure of BNNTs represented by a chiral angle. The motion transmission efficiency of the parallel junctions increases up to a maximum for certain BNNTs depending on displacement rates. It then decreases with increasing chiral angles. For cross junctions, the angular motion transmission ratio increases with decreasing chiral angles of the driven BNNTs, while the translational one exhibits the opposite trend.
\end{abstract}

\maketitle

\section{Introduction}
Gears are key components in most machines and devices with moving parts. The earliest example of gear dates from the 4th century BC, and is preserved at the Luoyang Museum in China.\cite{Needham1965} Gears are considered to be not only one of the greatest inventions of all time, but also essential for future nanotechnology. Top-down design has emerged as a major methodology for inventing nano-machines such as molecular analogues of cars,\cite{Rapenne2017} motors,\cite{Kassem2017} elevators,\cite{Badjic2004} shuttles\cite{Brouwer2001}, and so forth. However, the lack of an effective motion transmission system remains a critical problem for the top-down design of nano-machines inspired by their macroscopic-world counterparts. Difficulties include implementing well-positioned meshing teeth on nanostructures,\cite{Pantarotto2008} as well as strong friction and adhesion due to the extreme surface-to-volume ratio of nanomaterials.\cite{Kim2007} To this end, we present an idea to use the long-range interaction between atoms as meshing gear teeth for transmitting power, taking boron nitride nanotube (BNNTs) junctions as examples.

BNNTs are promising building blocks of nanoscale devices and machines owing to their peculiar structure, superior mechanical strength, high thermal conductivity and chemical stability.\cite{Zhang2017,Arenal2010} A single-walled BNNT consists of a single layer of B and N atoms arranged in a hexagonal \textit{h}-BN lattice. The layer can be rolled up in different directions with respect to the central axis of the BNNT leading to different chiralities, just like in carbon nanotubes (CNTs).\cite{Golberg2010} At the junctions between two BNNTs, the tubes are held together by both van der Waals (vdW) and electrostatic forces. These forces depend on the relative crystalline orientations of the tubes,\cite{Wang2019c} i.e., on the local stacking of layers.\cite{Marom2010} Indeed, non-covalent interactions between CNTs and substrates have long been known to strongly depend on the interfacial stacking sequence.\cite{Kolmogorov2004,Falvo2000,Chen2013,Sinclair2018} Based on this feature, screw-like motions have recently been reported for walls in concentric CNTs.\cite{Guerra2017,Cai2014} Barreiro \textit{et al.} observed directional motion of a gold nanoparticle on a CNT in scanning electron microscopy experiments.\cite{Barreiro2008}

BNNTs can be expected to present similar dependences on the stacking sequence owing to their crystalline similarity to CNTs. However, the optimal stacking sequences of successive \textit{h}-BN sheets is known to be different from those of graphene sheets.\cite{Gilbert2019,Wang2018} Moreover, the non-covalent force between BNNTs is much stronger than that between CNTs due to interlayer electrostatic interaction.\cite{Falin2017} Recently, ultrahigh friction has been measured between shells of multi-walled BNNTs in atomic force microscopy experiments.\cite{Nigues2014} To explore gear effects at the interface between \textit{h}-BN layers based on possible stacking sequence features, we conduct molecular mechanics simulations of non-covalent junctions between chiral single-walled BNNTs. Two different types of junctions are considered here: a parallel junction between two aligned BNNTs and a cross junction between two perpendicular ones.

\section{Methods}

\begin{figure}[htp]
\centerline{\includegraphics[width=8cm]{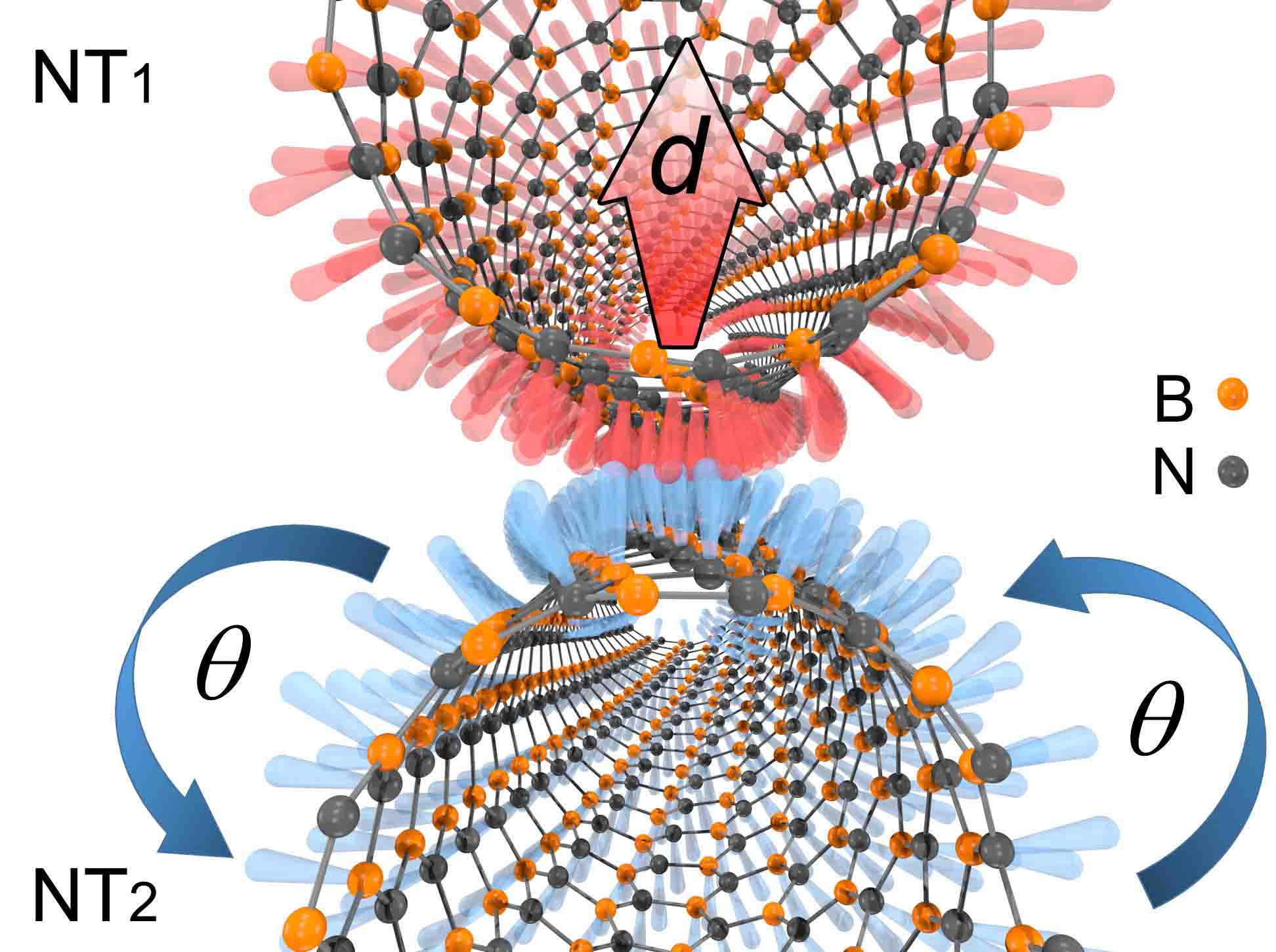}}
\caption{\label{F1}
Model setup for a parallel non-covalent junction between two BNNTs. The overlap between cones illustrates how the atoms interact with those from the other tube. The arrows indicate the direction of the motion of the BNNTs.}
\end{figure}

A wall of a BNNT can be described as a \textit{h}-BN sheet rolled up at a specific angle $\phi$, and thus characterized by a pair of integers $n$ and $m$ defining a vector in the \textit{h}-BN lattice.\cite{Arenal2010,Golberg2010,Zeng2010} This vector forms a circumference when the \textit{h}-BN sheet is rolled up to create a BNNT. The tube radius $R$ and the chiral angle $\phi$ are well-defined functions of $n$ and $m$, namely

\begin{equation}
\label{Eq22}
\left \{
\begin{array}{l}
\phi=\tan^{-1}\left( \frac{\sqrt{3}m}{2n+m} \right) \\
R=\frac{a}{2\pi}\sqrt{n^{2}+nm+m^{2}}
\end{array}\right.
\end{equation}

\noindent where $a$ is the lattice constant. $\phi$ varies between $0$ and $30^{\circ}$, covering the spectrum from zigzag $\left(n,0\right)$ to armchair $\left(n,n\right)$ BNNTs. A driving tube (denoted as NT$_{1}$) is first caused to slide along its axis, while the other, NT$_{2}$, can freely move in the plane normal to its axis, and its response to the displacement of NT$_{1}$ is simulated by a molecular mechanics procedure in which the equilibrium atomistic configuration of the atoms is computed at each iteration by minimizing the total potential energy of the system.\cite{Wang2009d,Wang2009c,Wang2010b,Wang2007,Wang2007a} The simulation scheme is also illustrated in an animation provided as part of the Supporting Information.

The potential energy of the system is given by the sum of those of covalent bonding (cov), long-rang van der Waals (vdW), and electrostatic (elec) interactions,

\begin{equation}
\label{Eq0}
\varepsilon=\sum \limits_{i=1}^{N_{1}-1}{\sum\limits_{\substack{j=i+1} }^{N_{1}}{ \varepsilon^{cov}_{ij}}}
                 +\sum \limits_{k=1}^{N_{2}-1}{\sum\limits_{\substack{l=k+1} }^{N_{2}}{ \varepsilon^{cov}_{kl}}}
								 +\sum \limits_{i=1}^{N_{1}}{\sum\limits_{\substack{k=1} }^{N_{2}}{ (\varepsilon^{vdW}_{ik}+\varepsilon^{elec}_{ik})}}
\end{equation}

\noindent where $N_{1}$ and $N_{2}$ are the total number of atoms in the NT$_{1}$ and NT$_{2}$, respectively. $\varepsilon^{cov}$ is given by the three-body Tersoff potential,\cite{Tersoff1988}

\begin{equation}
\label{Eq1}
\varepsilon^{cov}_{ij}=a_{ij} \left[\varphi^R\left(r_{ij}\right) + b_{ij}\varphi^A\left(r_{ij}\right)\right],
\end{equation}

\noindent where $\varphi^R$ and $\varphi^A$ denote the interatomic repulsion and attraction terms between the valence electrons, respectively. $a_{ij}$ is a scale factor depending on the interatomic distance $r_{ij}$. The many-body effects are included in the bond-order function $b_{ij}$, which depends on the interatomic distance, the bond angle, the dihedral angle and the bond conjugation. More details including the parameterization and benchmarks for this potential is provided elsewhere.\cite{Los2017,Tang2011}

The inter-tube electrostatic potential $\varepsilon^{elec}$ is calculated by a pairwise Coulombic function

\begin{equation}
\label{Eq2}
\varepsilon^{elec}_{ik} =C\frac{q_{i}q_{k}}{r_{ik}},
\end{equation}

\noindent where $C$ is the Coulomb's constant, $q=0.42$ and $-0.42$ as the effective partial charges for B and N atoms, respectively.  

The Lennard-Jones (LJ) force field is employed to describe $\varepsilon^{vdW}$,\cite{Thamwattana2007}

\begin{equation}
\label{Eq3}
\varepsilon^{vdW}_{ik} = 4 \epsilon \left[ \left( \frac{\sigma}{r_{ik}}  \right)^{12} -  \left( \frac{\sigma}{r_{ik}}  \right)^{6}  \right]
\end{equation}

\noindent with potential well depths of $\epsilon=4.116$, $5.085$ and $6.281\;\mathrm{meV}$ and equilibrium distances of $\sigma=0.3453$, $0.3409$ and $0.3365\;\mathrm{nm}$ for the B-B, B-N and N-N interactions, respectively. The long-range interaction cutoff radius is set to $1.8\;\mathrm{nm}$. The Kolmogorov-Crespi (KC) force field has been reported to provide an improved description to the interaction potential between atomic layers at a high load,\cite{Kolmogorov2005} and has recently been parametrized for flat boron nitride sheets.\cite{Maaravi2017} However, minor differences between the KC and LJ models are expected for these simulations, since the BNNTs are placed in vacuum with no external pressure. The inter-tube force should therefore mainly depend on the positions of the potential peaks, while the depth of the potential well should have a far weaker effect.\cite{Reguzzoni2012} Moreover, the use of the constant effective ionic charges is expected to hold as a reasonable approximation for our periodic system according to a comparison made by Maaravi \textit{et al.} \cite{Maaravi2017}. Note that we consider junctions between free-standing BNNTs without significant deformation of those adsorbed in a substrate.\cite{Zhao2014,Qi2018}

\section{Results and Discussion}
\subsection{Parallel Junctions}

\begin{figure}[htp]
\centerline{\includegraphics[width=8cm]{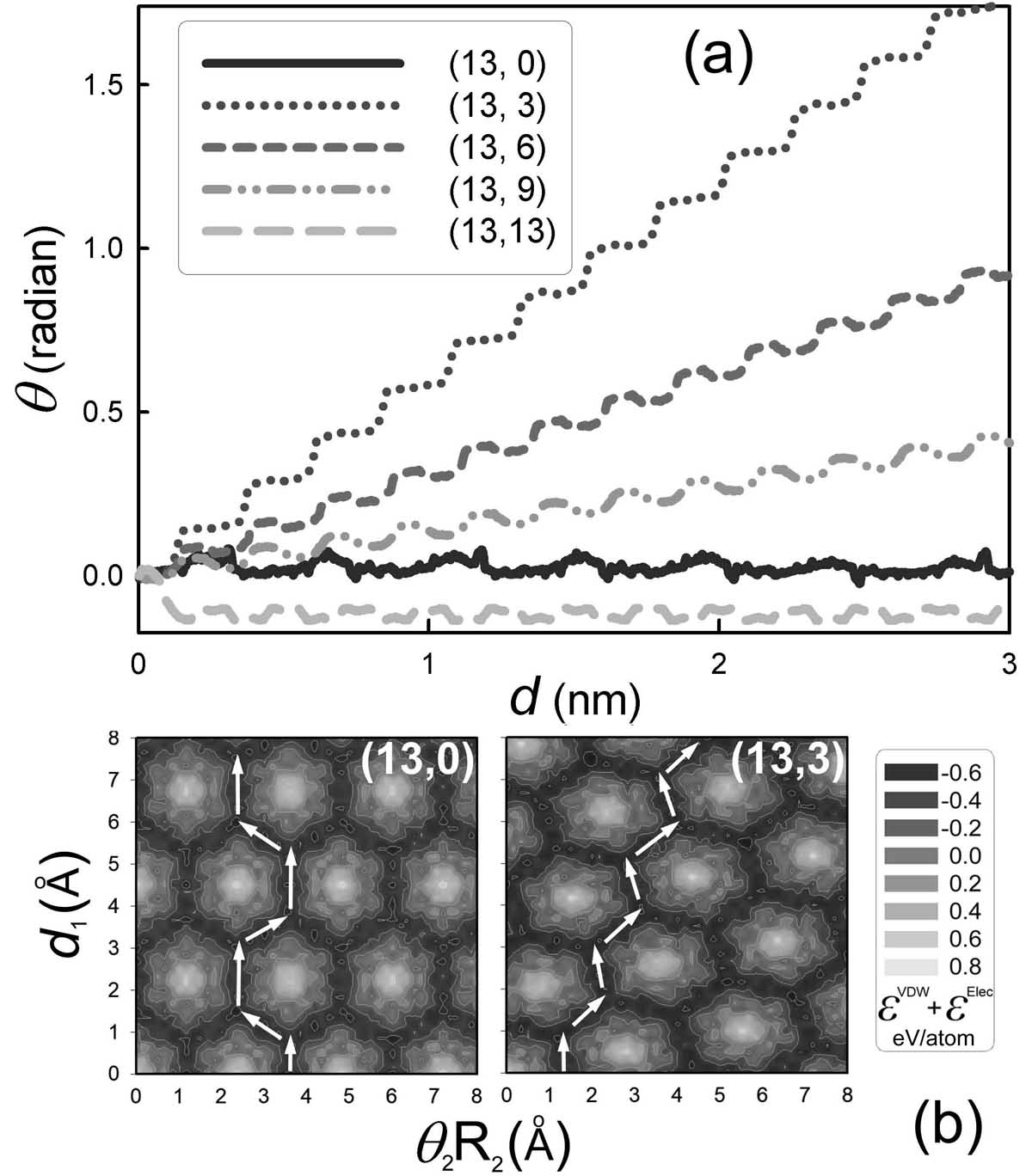}}
\caption{\label{F2}
(a) Rotation angle $\theta_{2}$ of NT$_{2}$ versus sliding distance $d_{1}$ of NT$_{1}$ for different parallel junctions. The displacement rate of NT$_{1}$ is \SI{5}{\angstrom} each $10^{6}$ iterations. (b) Profile of potential energy with respect to $d_{1}$ and $\theta_{2}$. The arrows indicate energetically optimized paths on BNNT surfaces.}
\end{figure}

We first consider a simple case of a parallel junction between two identical infinite $\left(n,m\right)$ BNNTs aligned side by side. NT$_{2}$ is observed to rotate around its axis in response to the translational motion of NT$_{1}$ as shown in Fig.~\ref{F2}(a). On average, the rotation angle $\theta_{2}$ of the NT$_{2}$ increases linearly with increasing displacement $d_{1}$ of NT$_{1}$, with a proportionality constant dependent on tube chirality. For instance, NT$_{2}$ rotates faster for BNNTs with small chiral angles, as seen in the zigzag (13,0) BNNT. However, the zigzag and armchair BNNTs oscillate back and forth instead of rotating continuously. Moreover, all these $\theta_{2}$ curves contain oscillations with a certain period, which corresponds to the dimension of a unit cell in the BNNT lattice. This typical oscillation is often measured by scanning probe microscopy experiments,\cite{Vilhena2016a,Dienwiebel2004} and is correlated with the distribution of the potential energy at the interface.  

The atoms in a BNNT are arranged in a hexagonal pattern, with their orbitals forming an eggbox-like potential landscape as shown in Fig.~\ref{F2}(b). Therefore, it stands to reason that, when two BNNTs are put in contact, there will be some specific directions in which the BNNTs can slide along each other more easily. Those easy directions are the keys to the chirality dependence of the motion transmission behavior. BNNTs of different chiralities result in different distributions of the potential energy at the surface,\cite{Verhoeven2004} which lead to different energetically optimized paths (EOPs). In Fig.~\ref{F2}(b), the rotation of the NT$_{2}$ is represented by the abscissas while the sliding of the NT$_{1}$ is given by the ordinates. It is clear that the shape and periodicity of the EOPs are consistent with those of the corresponding $\theta$-$d$ curves in Fig.~\ref{F2}(a). For instance, following the EOP, the angle $\theta_{2}$ of the $\left(13,0\right)$ BNNT oscillates back and forth when NT$_{1}$ slides (a displacement along the horizontal axis), a behavior in contrast to that of the $\left(13,3\right)$ BNNT. Note that a non-orthogonal EOP in the $R_{2}-d_{1}$ plane will results in a helical orbit on the BNNT surface similar to that previously reported for a gold nanoparticle inside CNTs.\cite{Schoen2006}

\begin{figure}[htp]
\centerline{\includegraphics[width=8cm]{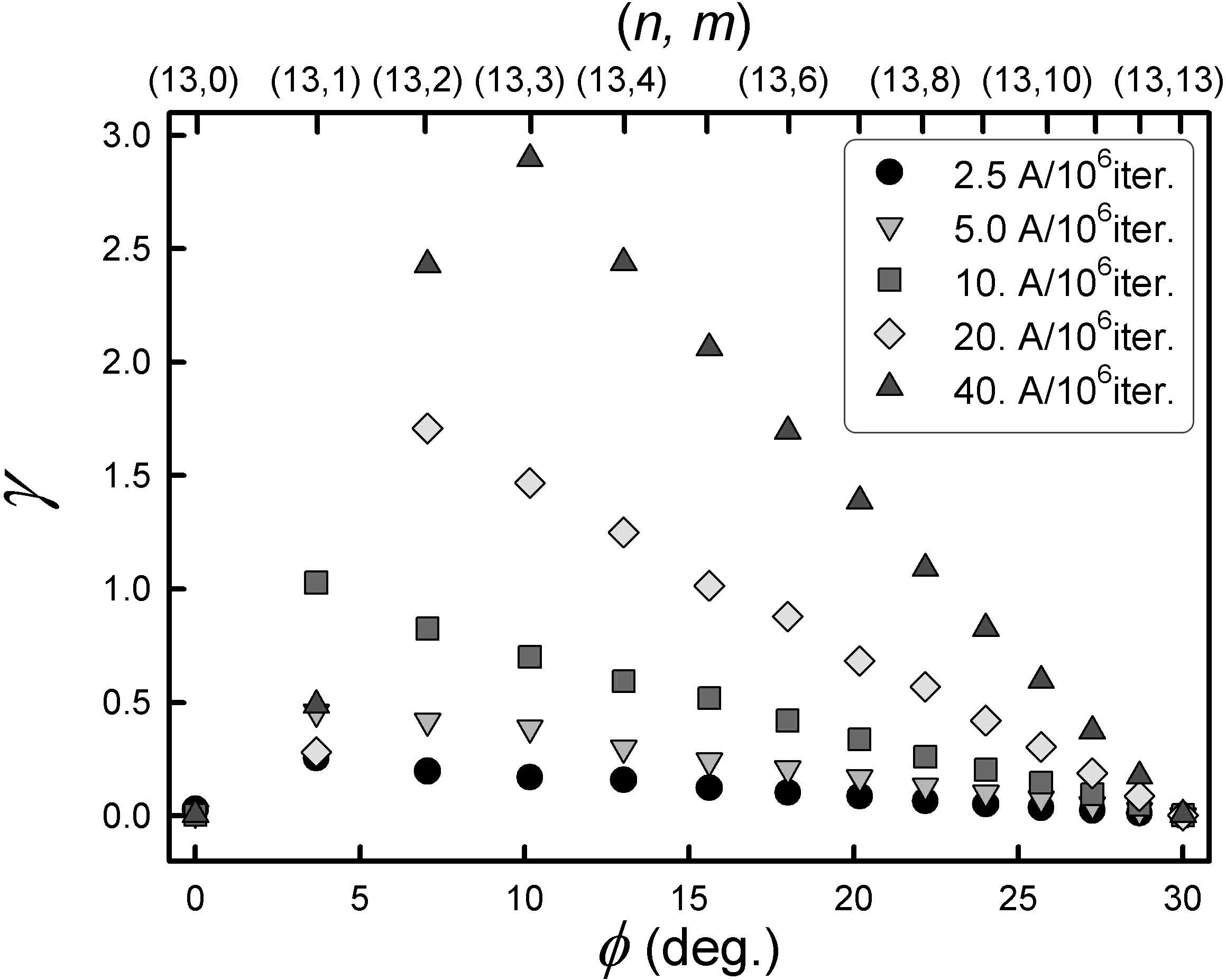}}
\caption{\label{F3}
Motion-transmission factor $\gamma$ (Eq.\ref{Eq4}) versus chiral angle $\phi$ for fourteen different $\left(13, m)\right)$ BNNTs with $m=0,1,2,...,13$. Different symbols stand for data from simulations performed for different rates of displacement of NT$_{1}$.
}
\end{figure}

A dimensionless factor proportional to the slope of the $\theta$-$d$ curves in Fig.~\ref{F2}(a) can be defined, to represent the efficiency of motion transmission of the parallel junctions between BNNTs,

\begin{equation}
\label{Eq4}
\gamma = \frac{R_{2}\theta_{2}}{d_{1}}.
\end{equation}

\noindent where $\theta$ is measured in radians. $\gamma$ is plotted as a function of the chiral angle $\phi$ in Fig.~\ref{F3}. It decreases linearly with increasing $\phi$, while it is almost zero for the zigzag and armchair BNNTs since they oscillate instead of rotating. $\gamma$ also increases with increasing sliding rates of NT$_{1}$. This rate dependence can be understood in terms of the inertia of the minimization procedure at branching points of the potential energy landscape where it has to choose between two energetically degenerate paths [see Fig. ~\ref{F2}(b)]. 

\begin{figure}[htp]
\centerline{\includegraphics[width=6cm]{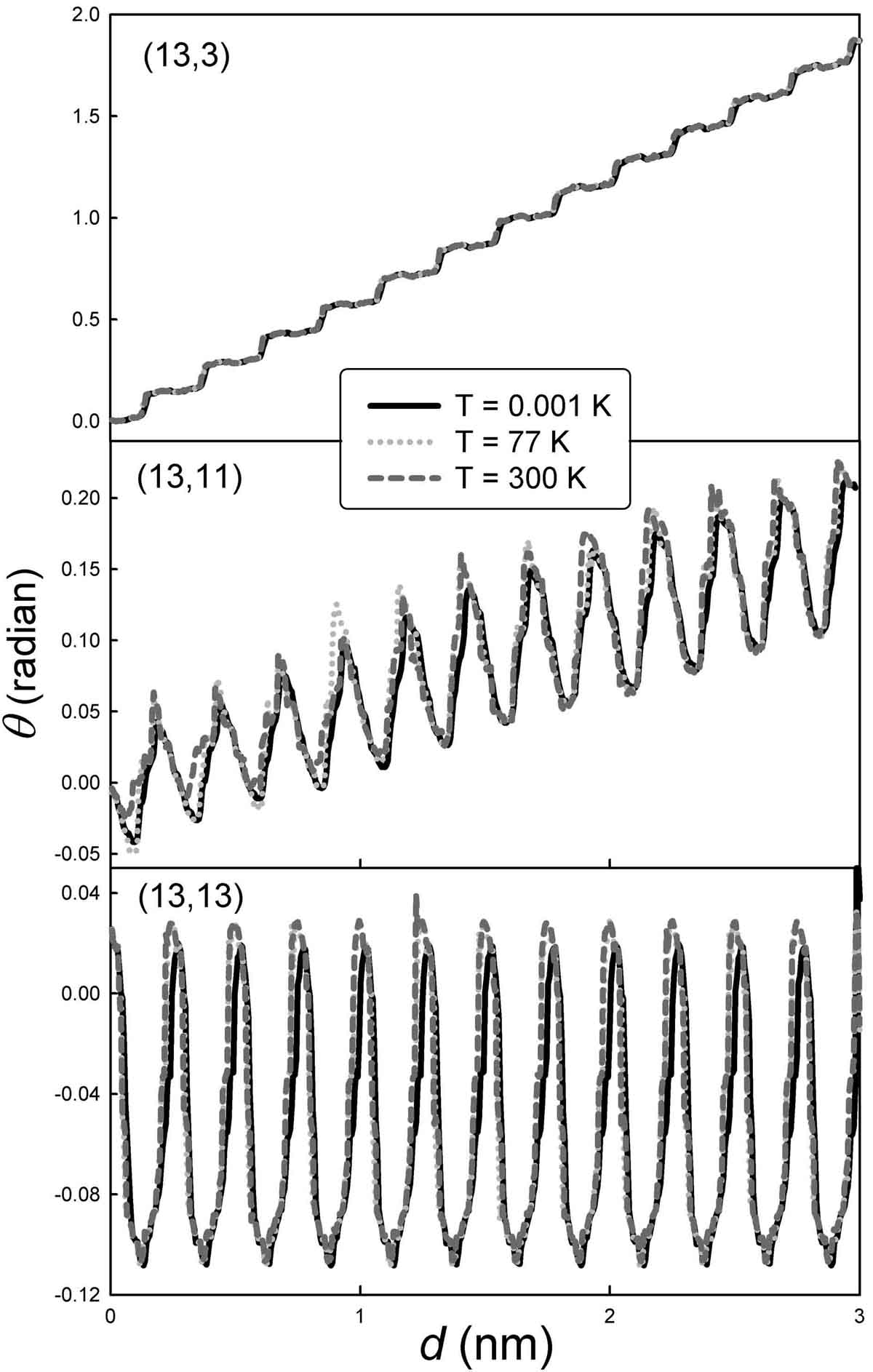}}
\caption{\label{F3b}
Rotation angle $\theta_{2}$ of NT$_{2}$ versus sliding distance $d_{1}$ of NT$_{1}$ calculated from MD simulations at three different temperatures for different pairs of BNNTs. The displacement rate of NT$_{1}$ is \SI{5}{\angstrom} per nanosecond.
}
\end{figure}

The functional relation between $\gamma$ and the displacement rate cannot be quantitatively predicted here since the energy-minimization simulation does not represent actual displacements as functions of time; however, the same qualitative phenomenon will appear in molecular dynamics (MD) because of the kinetic energy of the system \cite{Wang2011a,Guo2015}. I therefore perform MD simulations to check the effect of temperature on the motion transmission behavior of the BNNT by using the Nos\'{e}-Hoover thermostat with a time step of 1 femtosecond \cite{Plimpton1995,Wang2018,Wu2019,Li2013,Lin2014}. Fig.~\ref{F3b} shows the rotation angle of NT$_{2}$ when NT$_{2}$ is made to slide at three different temperatures, $0.001$, $77$ (liquid nitrogen temperature) and $300$ K. The comparison between the results at the three temperatures shows that the rotation angle remains almost the same for the tested pairs. However, it can be seen that the kinetic energy adds uncertainty to the displacement of BNNTs due to the random movement of the atoms that increases the probabilities of trajectories away from the potential energy minima. 

\begin{figure}[htp]
\centerline{\includegraphics[width=6cm]{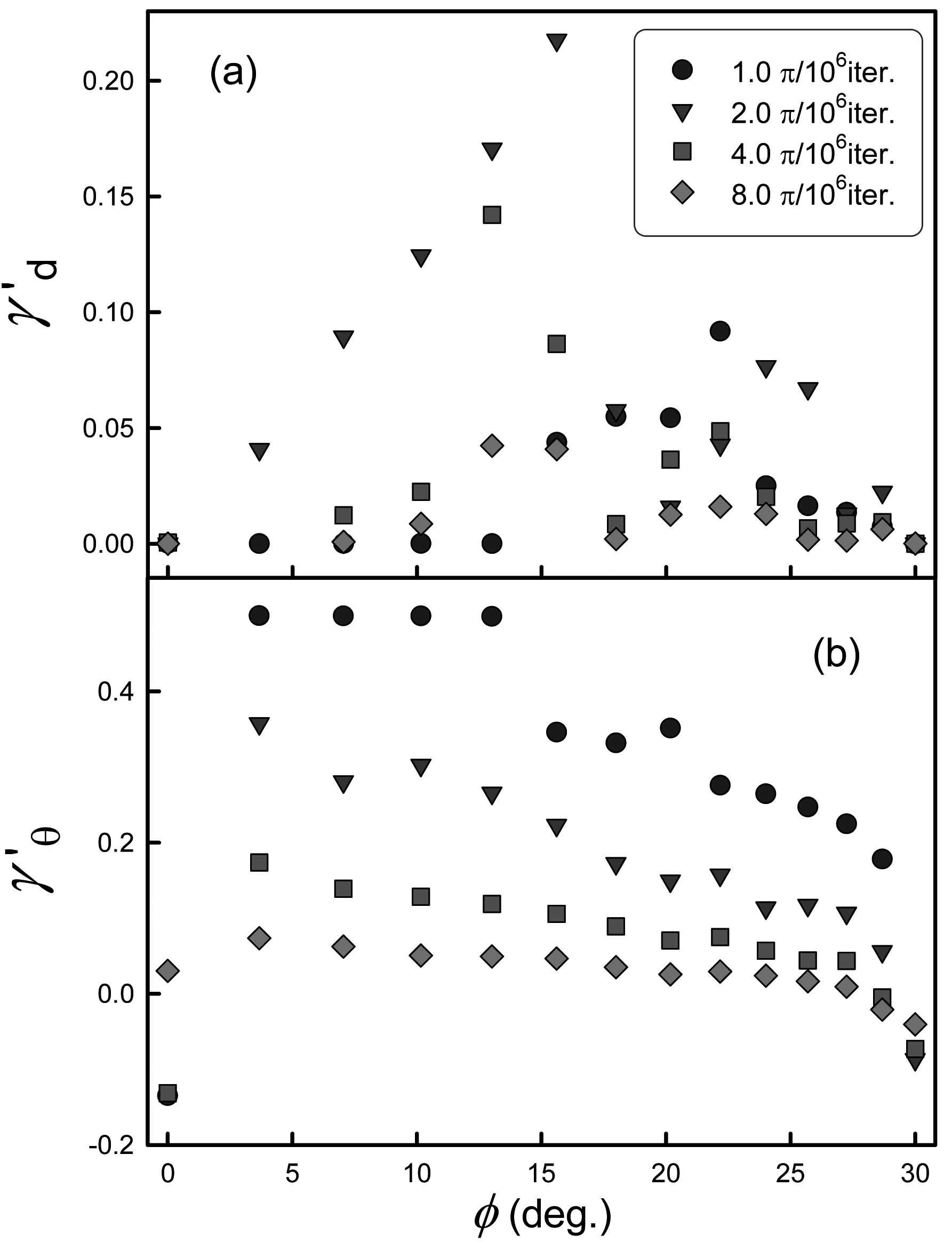}}
\caption{\label{F4}
Motion-transmission factors $\gamma'_{d}$ (a, Eq.\ref{Eq5}) and $\gamma'_{\theta}$ (b, Eq.\ref{Eq6}) versus $\phi$ at five different rotating rates of NT$_{1}$.
}
\end{figure}

It is found that the aforementioned motion transmission is reversible, in the sense that NT$_{2}$ will slide along its axis if the driving NT$_{1}$ rotates. NT$_{2}$ will also rotate at the same time. We find that the sliding distance $d_{2}$ and the rotation angle $\theta_{2}$ of NT$_{2}$ are well-defined functions of the rotation angle $\theta_{1}$ of NT$_{1}$. Two dimensionless factors can be defined to represent the efficiency of this motion transmission, 

\begin{equation}
\label{Eq5}
\gamma'_{d} = \frac{d_{2}}{\theta_{1}R_{1}}
\end{equation}

\noindent is the translational factor with $\theta$ measured in radians, and 

\begin{equation}
\label{Eq6}
\gamma'_{\theta} = \frac{\theta_{2}}{\theta_{1}}
\end{equation}

\noindent is the rotational factor. It can be seen from Fig.~\ref{F4} that $\gamma'_{d}$ increases with increasing $\phi$ and reaches a maximum at $\phi  \approx 15^{\circ} - 20^{\circ}$, before it decreases to about zero for the armchair BNNT. $\gamma'_{d}$ is almost zero at low rotation rates of the NT$_{1}$, it has a maximum value at an intermediate rotation rate, and decreases when the NT$_{1}$ rotates faster. $\gamma'_{\theta}$ decreases with increasing $\phi$ except for the zigzag BNNT, and increases with decreasing rotation rates of the NT$_{1}$.

\begin{figure}[htp]
\centerline{\includegraphics[width=6cm]{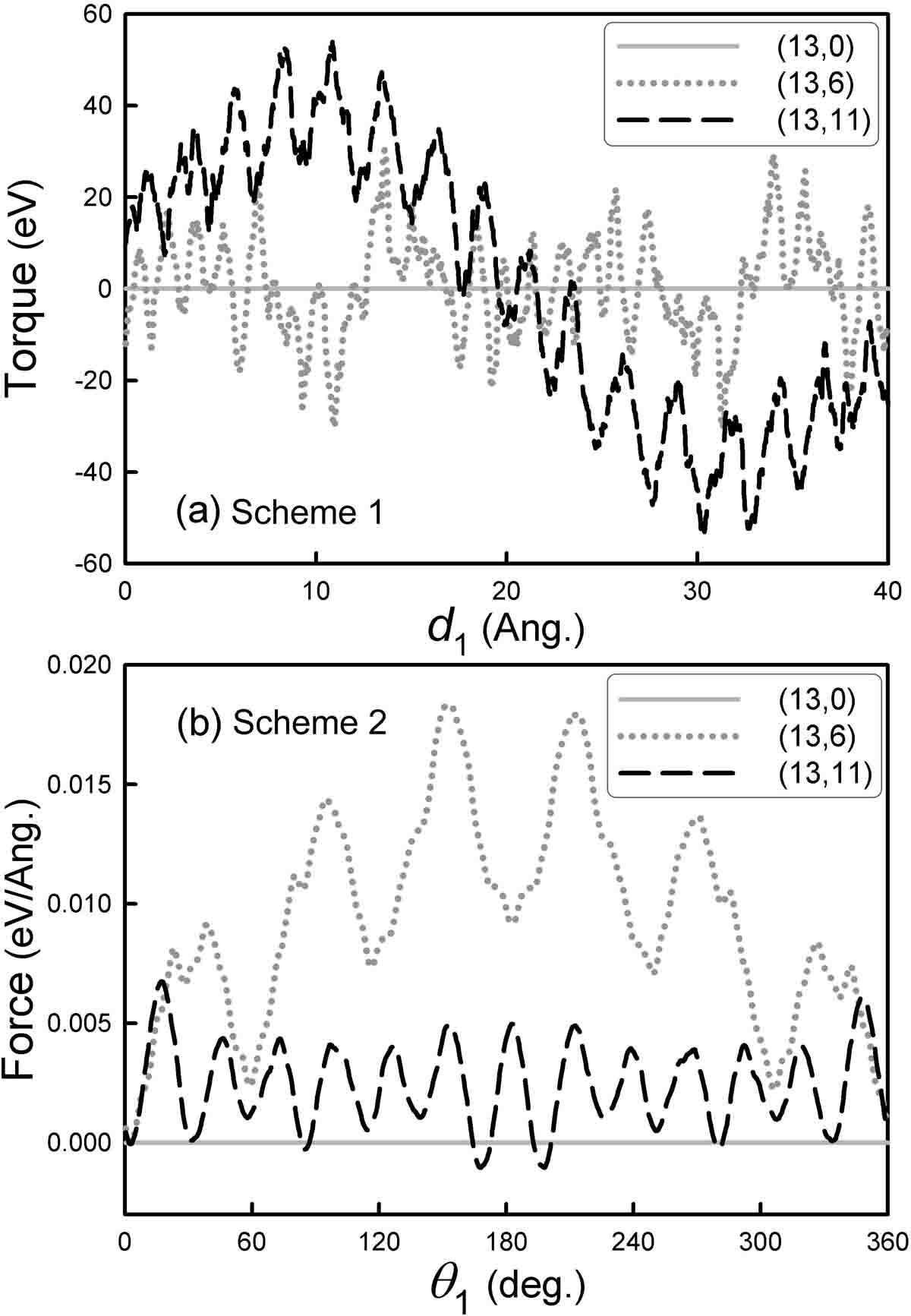}}
\caption{\label{F5}
Torque (a) or force (b) acting on NT$_{2}$ in the axial direction per angstrom of tube length when NT$_{1}$ slides along or rotates around its axis, respectively. NT$_{2}$ is fixed during the computation of the torque and the force, but remains free to move in all other simulations presented in this work.}
\end{figure}

An interesting question is how much mechanical load can be transmitted between the tubes in a parallel junction. The magnitude of the transferable torque and force depends on the specific boundary configuration. In the aforementioned simulations, the driven tube (NT$_{2}$) is free to move in the plane normal to its axis. Such a system can be used to transmit motion but cannot be used to transmit load since the inter-tube distance is adjusted spontaneously and the inter-tube normal stress tends to be zero. To transmit force and torque, the inter-tube distance needs to be fixed through the application of an external pressure between the tubes, just like when fixing the positions of the central axis of two gears before meshing their teeth to transmit power. To study this scenario, additional simulations are carried out with a fixed inter-tube distance of $\SI{3.1}{\angstrom}$. A rough estimation of the torque and the force is performed when NT$_{1}$ is made to slide [Fig.~\ref{F5}(a)] or to rotate [Fig.~\ref{F5}(b)]. The torque and the force acting on NT$_{2}$ vary with different BNNT types. It is observed that the torque and the force are almost zero for the zigzag $\left(13,0\right)$ BNNTs and are higher for the chiral ones, and that they reach up to \SI{40}{\eV} and \SI{0.02}{\eV\per\angstrom} per angstrom of tube length, respectively.

\subsection{Cross Junctions}

\begin{figure}[htp]
\centerline{\includegraphics[width=8cm]{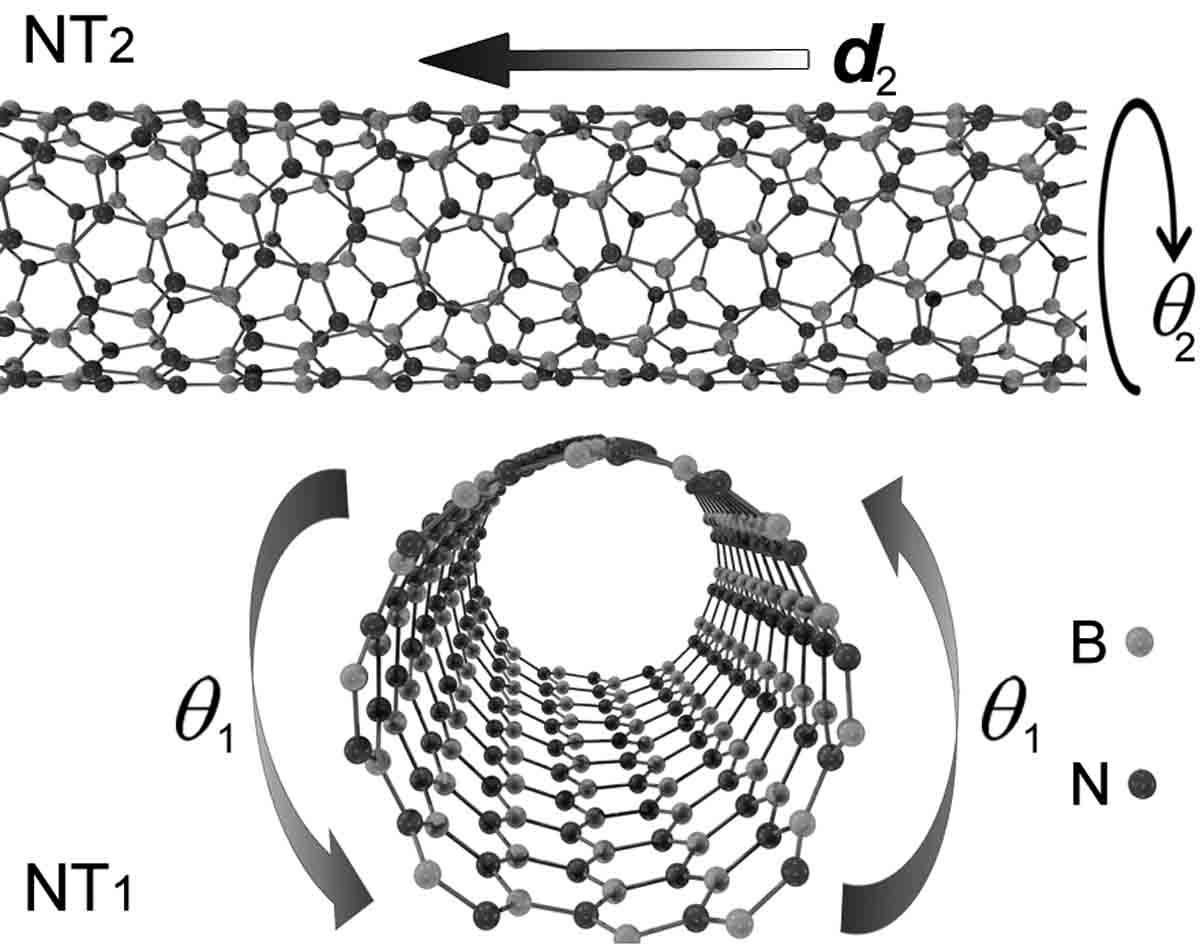}}
\caption{\label{F6}
Model setup for a BNNT cross junction. NT$_{1}$ is made to rotate around its axis by an angle $\theta_{1}$. NT$_{2}$ slides a distance $d_{2}$ and rotates by an angle $\theta_{2}$ in response.}
\end{figure}

The cross junction of BNNTs can also be used to transmit motion. As shown in Fig.~\ref{F6}, two BNNTs are placed perpendicular to each other in non-covalent interaction. NT$_{1}$ is then made to rotate around its axis in analogy to experiments \cite{Fennimore2003,Cohen-Karni2006} with a rotation rate of $2\pi\,$ per iteration of $10^{6}$ steps. The displacement $d_{2}$ and the rotation angle $\theta_{2}$ of the NT$_{2}$ are then measured as functions of $\theta_{1}$.

\begin{figure}[htp]
\centerline{\includegraphics[width=8cm]{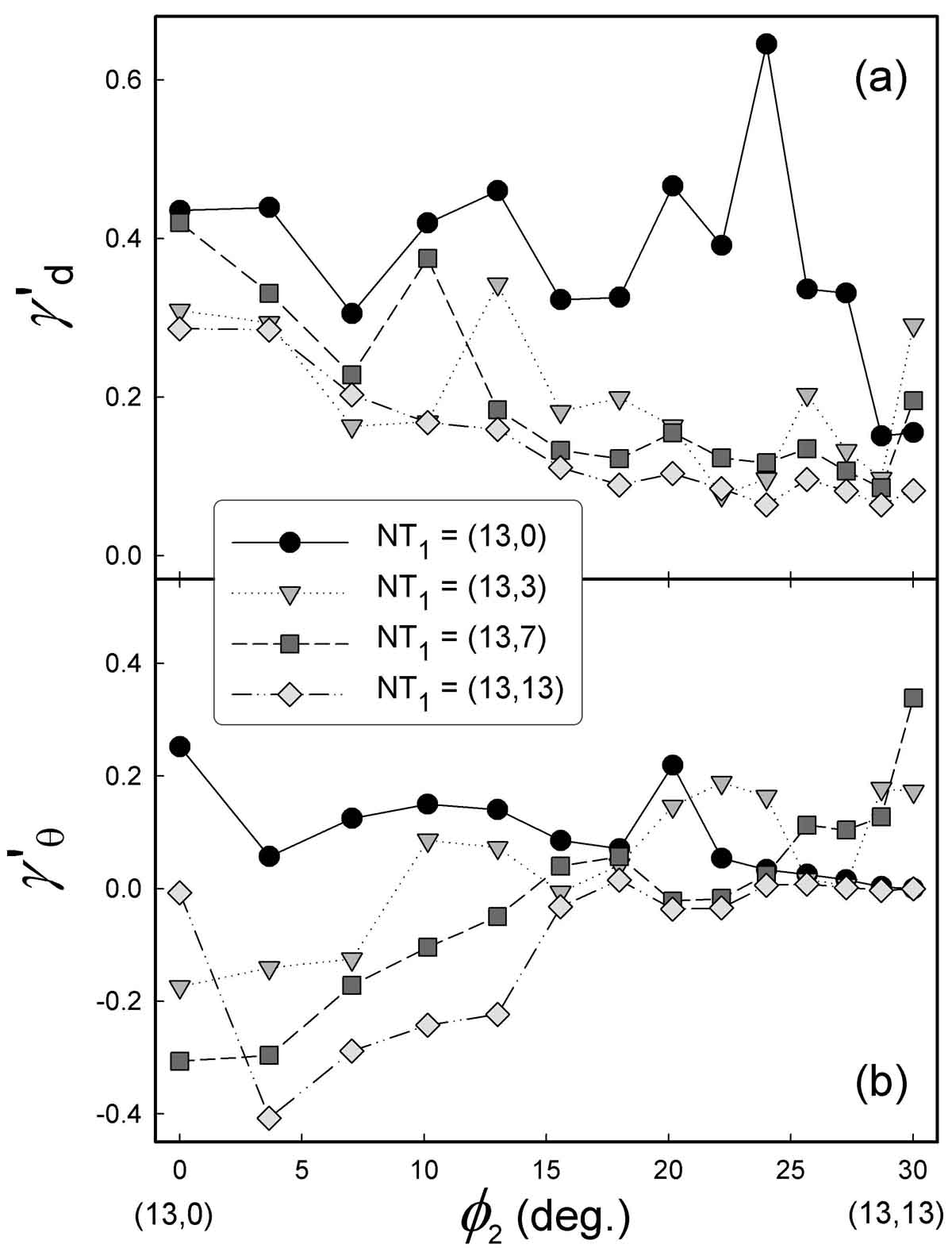}}
\caption{\label{F7}
  Transmission ratios $\gamma'_{d}$ (a) and $\gamma'_{\theta}$ (b) versus the chiral angle of the driven NT$_{2}$ for each driving $\left(13, m\right)$, with $m = 0,3,7,13$. }
\end{figure}

In Fig.~\ref{F7}, we plot $\gamma'_{d}$ and $\gamma'_{\theta}$ as functions of the chiral angle of the NT$_{2}$ for different possible NT$_{1}$. $\gamma'_{d}$ is found to decrease with increasing $\phi_{2}$ and is in general larger for the zigzag $\left(13,0\right)$ NT$_{1}$ and lower for the armchair $\left(13,13\right)$ one. Unlike $\gamma'_{d}$, which always has positive values, $\gamma'_{\theta}$ can be either positive (counterclockwise rotation) or negative (clockwise rotation) depending on the bi-chirality. $\gamma'_{\theta}$ is found to be negative for $\phi_{2}<20^{\circ}$ and to increase with increasing $\phi_{2}$, except for the zigzag tube. It deceases to about zero for the armchair NT$_{2}$. $\gamma'_{d}$ is higher for certain tube pairs with a preference for large $\left| \phi_{1}-\phi_{2} \right|$. The data are from simulations with $636$ pairs of BNNTs with different combinations of chiralities. A complete list of these bi-chiral BNNT pairs and their motion transmission factors are provided in the Supporting Information.

\section{Conclusion}
In conclusion, significant chirality-dependent gear effects are revealed at BNNT parallel and cross junctions. Atoms are used as gear teeth to transmit power orthogonally via long-range interactions. By this atomistic mechanism, the translational motion of a BNNT can be spontaneously translated into a rotation of the adjacent one, and \textit{viceversa}. At parallel junctions, the rotational motion transmission factor decreases linearly with the increasing chiral angle, while the translational one has a maximum at an intermediate value. The chirality dependence is attributable to different potential energy landscapes formed from different stacking sequences. The displacement rate also exhibits influence on the efficiency of the motion transmission. The magnitudes of the transmittable force and torque are estimated for parallel junctions. The transmission efficiency of BNNT cross junctions shows preference for tubes of a small size with a large bi-chirality. The atomistic gear mechanism has important implications for the design of mechanical power/speed transmission systems using nanostructured junctions.


\end{document}